\documentclass{article}

\usepackage{arxiv}

\usepackage[utf8]{inputenc} 
\usepackage[T1]{fontenc}    
\usepackage{hyperref}       
\usepackage{url}            
\usepackage{booktabs}       
\usepackage{amsfonts}       
\usepackage{nicefrac}       
\usepackage{microtype}      
\usepackage{lipsum}		
\usepackage{graphicx,float}
\usepackage{doi}
\usepackage{stackengine}
\usepackage{amssymb}
\usepackage{mathtools}
\usepackage{cancel}
\usepackage{cleveref}
\def\subrangle#1{\stackengine{5pt}{}{$\!\scriptstyle #1$}{U}{l}{F}{F}{L}}
\usepackage[sorting=none]{biblatex}
\title{Impact of Diffusion on synchronization pattern of epidemics in nonidentical metapopulation networks}

\date{} 					

\author{%
\href{https://orcid.org/0009-0006-3159-603X}{\includegraphics[scale=0.06]{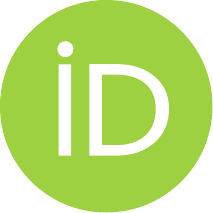}\hspace{1mm}Anika Roy}\textsuperscript{*}\\
International Institute of Information Technology \\
Hyderabad, India - 500032\\
\texttt{anika.roy@research.iiit.ac.in}
\And
\href{https://orcid.org/0009-0002-2024-7206}{\includegraphics[scale=0.06]{orcid.pdf}\hspace{1mm}Ujjwal Shekhar}\textsuperscript{*} \\
International Institute of Information Technology\\
Hyderabad, India - 500032\\
\texttt{ujjwal.shekhar@research.iiit.ac.in}
\And
\href{https://orcid.org/0000-0002-3036-8223}{\includegraphics[scale=0.06]{orcid.pdf}\hspace{1mm}Aditi Bose} \\
International Institute of Information Technology\\
Hyderabad, India - 500032\\
\texttt{aditi.bose@research.iiit.ac.in}
\And
\href{https://orcid.org/0000-0003-4043-1422}{\includegraphics[scale=0.06]{orcid.pdf}\hspace{1mm}Subrata Ghosh} \\
International Institute of Information Technology\\
Hyderabad, India - 500082\\
\texttt{subrata.ghosh@research.iiit.ac.in}
\And
\href{https://orcid.org/0000-0003-3194-7953}{\includegraphics[scale=0.06]{orcid.pdf}\hspace{1mm}Santosh Nannuru} \\
International Institute of Information Technology\\
Hyderabad, India - 500032\\
\texttt{santosh.nannuru@iiit.ac.in}
\And
\href{https://orcid.org/0000-0003-4165-3852}{\includegraphics[scale=0.06]{orcid.pdf}\hspace{1mm}Syamal Kumar Dana} \\
Department of Mathematics, Jadavpur University \\ Kolkata 700032, India\\
Division of Dynamics, Lodz University of Technology \\ Stefanowskiego 1/15, 90-924 Lodz, Poland\\
\texttt{syamaldana@gmail.com}
\And
\href{https://orcid.org/0000-0003-1971-089X}{\includegraphics[scale=0.06]{orcid.pdf}\hspace{1mm}Chittaranjan Hens} \\
Centre for CNS and Bioinformatics\\
International Institute of Information Technology\\
Hyderabad, India - 500032\\
\texttt{chittaranjan.hens@iiit.ac.in}
}
\renewcommand{\headeright}{}
\renewcommand{\undertitle}{}
\renewcommand{\shorttitle}{Impact of Diffusive Migration on Metapopulation Model Dynamics}

\hypersetup{
pdftitle={Impact of Diffusive Migration on Metapopulation Model Dynamics},
pdfsubject={q-bio.NC, q-bio.QM},
pdfauthor={Anika Roy, Ujjwal Shekhar, Aditi Bose, Chittaranjan Hens,  Santosh Nannuru, Syamal Kumar Dana,  Subrata Ghosh},
pdfkeywords={Synchronisation, Epidemic Models, Network Science}
}

\begin{document}
\maketitle
\footnotetext[1]{U.S. and A.R. have contributed equally to the work.}
\begin{abstract}
In a prior study, a novel deterministic compartmental model known as the SEIHRK model was introduced, shedding light on the pivotal role of test kits as an intervention strategy for mitigating epidemics. Particularly in heterogeneous networks, it was empirically demonstrated that strategically distributing a limited number of test kits among nodes with higher degrees substantially diminishes the outbreak size. The network's dynamics were explored under varying values of infection rate. In this research, we expand upon these findings to investigate the influence of migration on infection dynamics within distinct communities of the network. Notably, we observe that nodes equipped with test kits and those without tend to segregate into two separate clusters when coupling strength is low, but beyond a critical threshold coupling coefficient, they coalesce into a unified cluster. Building on this clustering phenomenon, we develop a reduced equation model and rigorously validate its accuracy through comprehensive simulations. We show that this property is observed in both complete and random graphs.
\end{abstract}


\section{Introduction}

Over the past two decades, there has been a notable increase in various infectious diseases within human populations. Examples include the Severe Acute Respiratory Syndrome (SARS) outbreak in 2003~\cite{anderson2004epidemiology,dye2003modeling}, the swine flu pandemic in 2009~\cite{coburn2009modeling,brockmann2013hidden,jamieson2009h1n1}, and the ongoing COVID-19 pandemic~\cite{maier2020effective,ciotti2020covid,ndwandwe2021covid}. They strain healthcare systems~\cite{van2023attacks,preti2020psychological}, causing a surge in patients and potentially high mortality rates~\cite{akin2020understanding,simonsen1998pandemic}. { Thus, suitable control (pharmaceutical and nonpharmaceutical) strategies are required to mitigate the severity of the outbreaks ~\cite{paper1, vespignani2, wang2016statistical}.
Against this backdrop, to acquire a better understanding of the disease transmission process and to assess the efficacy of intervention methods, researchers use compartmental mathematical models ranging from stochastic (Markovian or non-Markovian) to deterministic frameworks ~\cite{prx2020arena, paper1, vespignani2, Allen2008, WANG20161}.  In particular, computational epidemiological modelling has proven to be a significant tool for predicting COVID virus onset and spread ~\cite{paper18, paper17}. 
Note that, compartmental models, such as the SIR (Susceptible-Infectious-Recovered) model, are useful for understanding disease transmission in well-mixed populations. However, real-world interactions frequently occur in complex networks in which individuals are linked in a more detailed structure, and they can be effectively mapped with the real-world scenarios of disease transmission from a small to a large population scale.
Thus, both the modeling techniques and the structural properties of complex networks are essential for developing effective strategies to control the spread of epidemics. This interdisciplinary approach allows researchers to consider the nuances of real-world interactions and devise more realistic and impactful public health interventions ~\cite{pastor2001epidemic,pastor2015epidemic,pastor2002epidemic,hens2019spatiotemporal,brockmann2013hidden,belik2011natural,iannelli2017effective}.
For instance, complex networks play a crucial role in determining the optimal vaccination strategy~\cite{wang2016statistical,wang2017vaccination}. Also, several structural properties of complex networks are used as guidelines for shaping up the outbreak distribution~\cite{hindes2022outbreak} and may provide suitable paths of disease extinction~\cite{holme2013extinction,hindes2016epidemic}. Both networks and underlying disease dynamics are extensively used for predicting the epidemic propagation patterns~\cite{hens2019spatiotemporal,iannelli2017effective}, and for identifying super-spreading events~\cite{james2007event,nielsen2021covid} among metacommunities. The general dynamics that capture the states of compartments in a network of metacummunities can be written as  $\bf{{\dot{x}_i}={ F(x_i)}+ \epsilon \sum_j A_{ij} G(x_i,x_j)}$,
where  the vector ${\bf x_i}$ represents the current state of each compartment of the $i$ th community, and $\epsilon$  determines the strength of the diffusion 
(${\bf G(x_i,x_j) =x_j-x_i}$). Here, we investigate the strength of migration (diffusion constant $\epsilon$) in a network of metacommunies.
\par
The concept of migration highlights the significance of travel in the transmission of infections. With advancements in aviation and transportation networks, viruses can now spread more quickly than before \cite{paper16, paper15}.
From a theoretical perspective, these reaction-diffusion network dynamics are highly beneficial for comprehending the transmission of epidemics via human migration and predicting the arrival time~\cite{brockmann2013hidden} and velocity of the epidemic wave front~\cite{belik2011natural}.  In this context, the central focus of our research is to examine how epidemics spread and synchronize within complex networks, with the aim of revealing their spreading patterns.
In nonlinear oscillators having diffusive coupling, it is already established that after a critical diffusion, the coupled systems may follow the same trajectory, i.e update themselves synchronously \cite{boccaletti2002synchronization,arenas2008synchronization,kapitaniak2012synchronization}. This phenomenon is common in a wide range of coupled systems, from coordinated flight of birds~\cite{nagy2018synchronization} to rhythmic flashes of fireflies~\cite{ramirez2019modeling}, 
The synchronized beats of the heart~\cite{ivanov2009maternal}, firing of neurons in the brain~\cite{hopfield1995rapid}, and organized migration of animals~\cite{cote2017behavioural} are other examples where the exchange of signals between units plays a crucial role in shaping the synchronization dynamics. However, the role of diffusion in the synchronization pattern of epidemics has not been explored rigorously. Also, in the presence of non-identical meta communities (assume few communities have a lower infection rate), no one ever studied 
clustered synchronization of disease dynamics, except a recent study indicates that the interplay between two diseases, seasonal influenza and COVID-19, can affect their behaviors. Such coupling leads to the synchronization of both infections, with a more significant impact observed on the dynamics of influenza~\cite{paper3}. \par
Here,} to investigate the clustering of communities in terms of disease spread, we employ a complex six-compartment meta-population epidemiological model that includes the use of test kits, introduced in~\cite{paper1}. We explore the effect of diffusion coupling 
($\epsilon$) 
within the meta-population model and derive the critical coupling strength ($\epsilon_c$) of cluster and global synchronization. The rationale for selecting this model pertains to the necessity of delineating two distinct categories of nodes, distinguishing between those possessing diagnostic test kits and those lacking testing equipment. Based on this, it is discerned that nodes furnished with test kits and those devoid of test kits exhibit a tendency to segregate into discrete clusters at low coupling strength. However, upon surpassing the critical threshold in the coupling strength, all nodes amalgamate into a singular global cluster. 
\begin{figure}[htp!]
    \centerline{\includegraphics[width=1\columnwidth]{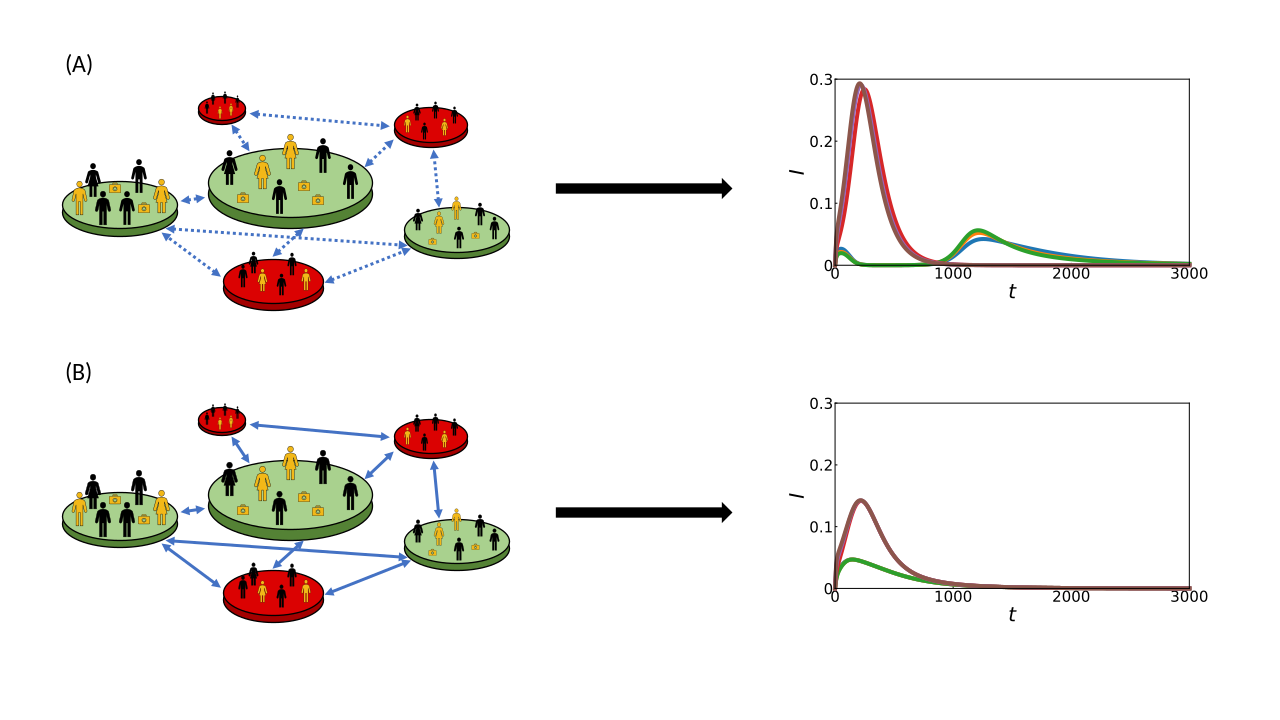}}
    \caption{An illustrative overview of our study depicts a meta-population model wherein each community shares an identical population size. Distinguished by colour, some communities possess access to test kits (depicted in green), while others lack this resource (depicted in red). Within each community, a subset of individuals are infected. In Case A, the network exhibits low coupling strength (represented by dotted arrows) between communities, while Case B portrays a scenario with increased coupling strength. In Case B, the infection trends of the communities cluster, forming distinct clusters under the influence of higher coupling strength.}
    \label{fig: basic_pic}
\end{figure}
Note that numerous synchronization patterns are conceivable, but the primary focus of our investigation pertains to the examination of partial or cluster synchronization~\cite{khanra2022identifying,arenas2008synchronization,   lahav2022topological,khanra2018explosive,kundu2017transition,mishra2023chimeras,pecora2014cluster,sorrentino2016complete,jalan2016cluster,khanra2022endowing, rakshit2023predicting, ghosh2020emergence}. 
In ecological contexts, cluster synchronization is used ~\cite{paper4, paper20, 10.1093/nsr/nwaa269, saha2020infection} to investigate how the connectivity of a dispersal network and varying coupling strengths affect the persistence of species in different patches or habitats.
In separate studies of bursting in neurons~\cite{paper5,ghosh2020emergence,saha2predicting,hens2015bursting}, researchers illustrate that a network of heterogeneous de-synchronized neurons, when subject to diverse memory settings, can ultimately lead to bursting and the formation of cluster synchronization beyond a certain coupling strength threshold. 
By building upon these existing works and contributing new insights into infection control strategies, our study aims to enhance our understanding of disease dynamics and inform the development of more effective public health interventions.
\par Our research specifically hones in on infection trends within interconnected communities, taking into account human mobility patterns driven by diffusion-like migration. 
Figure~\ref{fig: basic_pic} illustrates our research motivation. As inter-community migration strength increases, nodes segregate based on the presence or absence of test kits. This clustering becomes more pronounced with higher diffusion rates.
Additionally, the migration term's influence on the behavior of $I_{\text{max}}$ (peak of the infection curve) becomes apparent as diffusion strength changes. This influence stems from individuals in a compartment diffusing from nodes with higher population to those with lower population. Testkit communities, experiencing rapid hospitalization, have fewer individuals in the infected compartment, resulting in a higher influx of infected individuals from non-testkit to testkit communities. Thus, with increasing diffusion strength, infection peaks for non-testkit nodes decrease and testkit nodes increase as seen in the figure.

{ In this paper, we examine the process of cluster formation among nodes possessing test kits and those lacking test kits. We describe the model and quantities we work with in Sec. ~\ref{sec: 2}.  The detailed analysis for the effect of coupling strength in a complete graph is presented in Sec. ~\ref{sec: 3}. We analyse the
peak of infection ($I$\textsubscript{max}) for nodes with and without test kits, along with their respective averages, as a function of coupling strength ($\epsilon)$, for different values of initial fractions of nodes ($p$) equipped with test kits. We also demonstrated the error of global, cluster (with test kits) and cluster ( without test kits) synchronization in relation to coupling strength. 
In addition to our inter-cluster observations, our analysis unveils intriguing phenomena, including the emergence of two peaks in infection plots and an unexpected dip in $I$\textsubscript{max} due to the inherent characteristics of our model.{ To gain deeper insights, we have studied the phase plot of synchronization errors (global and cluster-wise) in ($\mathcal{R}_{0}$, $\epsilon$) and ($p$,$\epsilon$) plane. Here, $\mathcal{R}_{0}$ represents the basic reproduction number. Evidently, we construct a reduced equation~\cite{ghosh2020emergence} model and employ it to compute the thresholds for critical coupling strength (Sec. ~\ref{sec: 4}). Furthermore, we corroborated these findings by applying various network topologies, including regular graphs(Sec. ~\ref{sec: 3}) and Erd\H{o}s-R\'enyi network(Sec. ~\ref{sec: 5}). Finally, conclusive remarks are made with the results and key takeaways in Sec. ~\ref{sec: 6}.} 





\section{A brief description of the model }
\label{sec: 2}
We employ an epidemiological model, as detailed in a prior study~\cite{paper1}. This model extends the standard S-E-I-R framework by introducing a `hospitalized' compartment ($H$) to account for individuals requiring hospitalization due to the disease. Additionally, it incorporates a novel state variable, $K$, representing the quantity of available test kits. The motivation behind this extension lies in recognizing the pivotal role of test kits in mitigating the disease's spread. The model comprises five distinct compartments: Susceptible ($S$), Exposed ($E$), Infected ($I$), Recovered ($R$), and Hospitalized ($H$), each reflecting different stages of disease exposure and progression. Specifically, individuals transition from the susceptible to the exposed state through close contact with infected individuals, a process governed by the infection transmission rate $\beta$. The latent period and recovery rate are denoted by $\sigma$ and $\gamma$, respectively.
Notably, the transition of infected individuals to the hospitalized compartment depends on the availability of test kits and is expressed as a function of $K$, denoted as $\alpha(K)$. We posit that the rate of hospitalization is influenced by the number of available test kits and is modelled as a linear function, $\alpha(K) = \left(\alpha_0 + g_n(K)\right)$. Here, $\alpha_0$ represents the baseline hospitalization rate in the absence of any test kit.
The production of test kits is tied to the total infected population, with a production rate $\zeta$, while the efficacy of these kits diminishes due to production faults at a rate $\chi$. Furthermore, the distribution of test kits across communities is determined by $g_n(K) = \alpha_1 \times p_nK$, where $\alpha_1$ signifies the effectiveness of the test kit and $p_n$ is calculated as the fraction of test kits being allocated to a particular community $n$ out of the $l$ communities with access to test-kits, here, $p_n = \frac{d_n}{\sum_{i=1}^{l}d_i}$ while $d_n$ corresponds to the degree of a specific patch, where $n \in {1, 2, .., M}$. $\alpha_1$ is chosen keeping $\alpha(K)$ bounded; e.g., the diagnosis rate remains less than or equal to 1.

In the context of heterogeneous networks, the model's dynamics are coupled, resulting in a system of interconnected equations enumerated as follows:
\begin{align}
    \frac{dS_n}{dt} &= -\beta\left(\frac{S_nI_n}{N_n}\right) + \frac{\epsilon}{d_n}\displaystyle\sum_{m=1}^{M}{A_{mn}(S_m - S_n)} \\
    \frac{dE_n}{dt} &= \beta\left(\frac{S_nI_n}{N_n}\right)- \sigma E_n + \frac{\epsilon}{d_n}\displaystyle\sum_{m=1}^{M}{A_{mn}(E_m - E_n)}\\ 
    \frac{dI_n}{dt} &= \sigma E_n - \left(\alpha_0 + g_n(K)\right)I_n+ \frac{\epsilon}{d_n}\displaystyle\sum_{m=1}^{M}{A_{mn}(I_m - I_n)}\\ 
    \frac{dR_n}{dt} &= \gamma H_n + \frac{\epsilon}{d_n}\displaystyle\sum_{m=1}^{M}{A_{mn}(R_m - R_n)}\\ 
    \frac{dH_n}{dt} &= \left(\alpha_0 + g_n(K)\right)I_n - \gamma H_n\\ 
    \frac{dK}{dt} &= \zeta\displaystyle\sum_{m=1}^{M}{I_m} - \chi K. 
\end{align}
In this context, the unweighted adjacency matrix element $A_{mn}$ represents the interconnectivity between $M$ distinct communities, and $\epsilon$ denotes the strength of migration. The model accounts for migration dynamics through the diffusive term $\sum_{m=1}^{M}{A_{mn}(X_m - X_n)}$, which governs the exchange of individuals between these communities. Two critical model assumptions prevail: first, that the disease under examination does not result in any fatalities, and second, that the timescale of birth and death processes greatly exceeds that of the disease outbreak, rendering these demographic processes negligible and thus disregarded within the model. 
Before delving into the impact assessment of the coupling strength, it is also essential to establish a standard mathematical formulation for quantifying clustering. The following section outlines our formulation of the Synchronization Error, $E$.
We define synchronization errors at both global and cluster levels to quantitatively assess clustering~\cite{khanra2022endowing,khanra2022identifying,sorrentino2016complete}. The Global Synchronization Error is denoted as $E_{\mathrm{global}}$, while the cluster synchronization error is denoted as $E_{\mathrm{cluster}}$, abbreviated as GSE and CSE respectively.

This formulation of synchronization error employs a well-established mathematical concept, asserting that when a series of values are closely clustered, they are also close to their mean.

\begin{equation}
    E_{\mathrm{global}} = \Big\langle \sqrt{\langle (x_i-\bar{x})^2 \rangle\subrangle{\mathcal{N}}}\Big\rangle_{t}
\end{equation}

\begin{equation}
    E_{\mathrm{cluster}} = \left\langle \sqrt{\Big ({\frac{1}{\mathcal{N}_{k}} \sum_{i\in v_{k}} (x_i-\bar{x}_{k})^2 \Big )}}\right\rangle_{t}
\end{equation}

Here, $x_i$ could represent any compartment of the community, however, in our analysis, $x_i$ denotes the number of infected individuals in the $i^{th}$ community, $\bar{x}$ denotes the mean of the entire dataset, $\bar{x}_{k}$ is the mean of cluster $k$, and $\mathcal{N}_{k}$ is the size of cluster $k$.

\section{Impact of Coupling Strength on a Globally Connected Network}
\label{sec: 3}
Our investigation begins by examining a compact network comprising 20 fully interconnected communities, among which only 5 have access to test kits. Through simulations employing the Ordinary Differential Equations (ODEs) specified by the \textit{SEIHRK} model over a span of 3000 time steps, we observe two levels of synchronization in infection trends across communities. The first level of synchronization manifests within the test-kit and non-test-kit communities, demonstrating consistency in their infection trends. The second level is a global synchronization of infection trends across all communities at high coupling strengths. 
Given the inherent coupling of equations governing all compartments, synchronization in one compartment naturally extends to the others.

\begin{figure}[H]
    \centerline{\includegraphics[width=1\columnwidth]{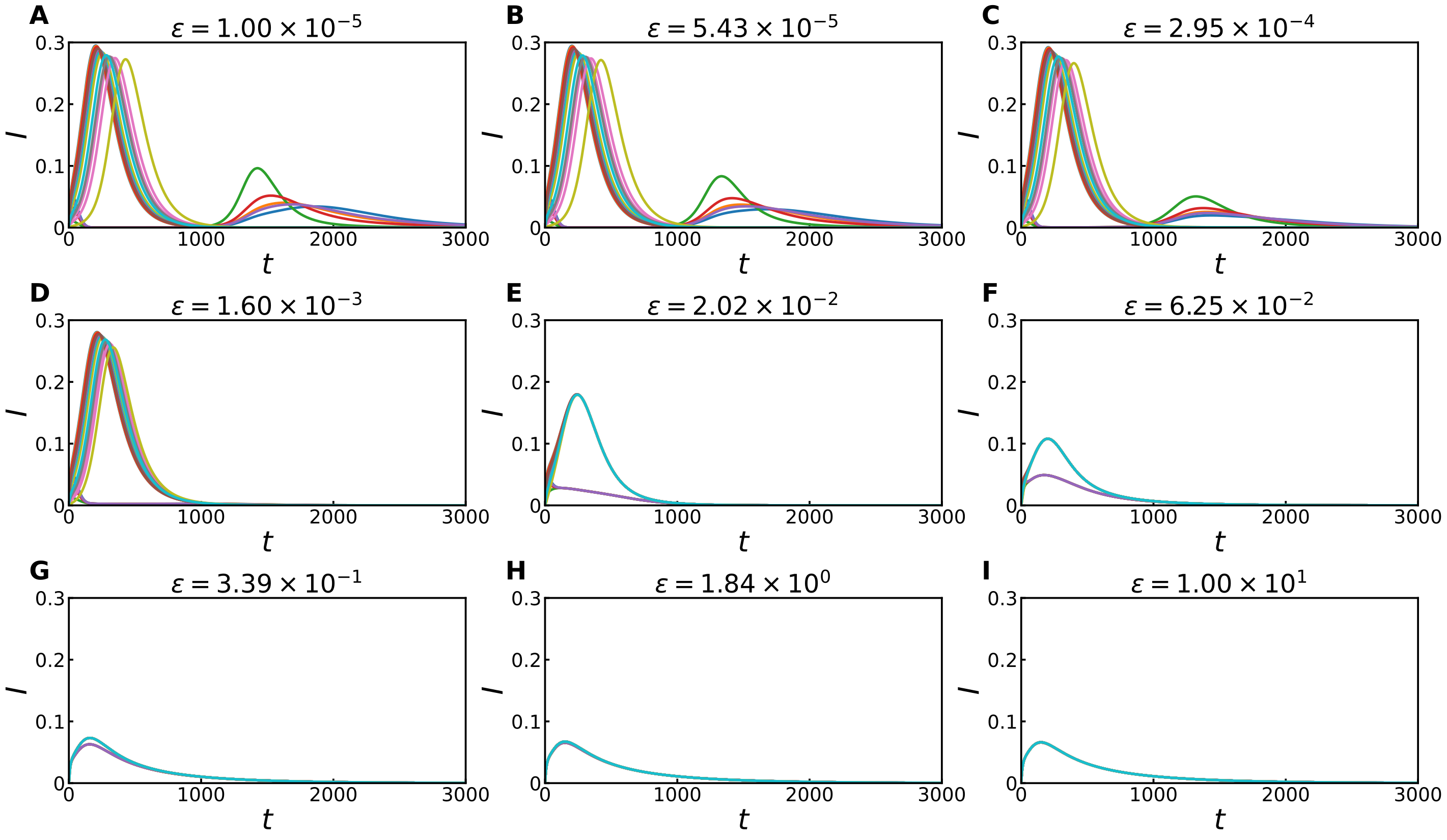}}
    \caption{Figures \textbf{A} through \textbf{I} show snapshots of infection trends among different communities with changing coupling strengths. There are $20$ communities, with $5$ having access to test kits. Each community has a population of $10000$ people, and $E_0$ for each community is randomly sampled between [1, 750]. The plot parameters used were $\sigma = 0.1,\alpha_0 = 0.01,\ \alpha_1 = 0.0001,\ \gamma = 0.07,\ \zeta = 0.02,\ \chi = 0.1$ and $\beta = 0.03$.}
    \label{fig: gradual_clustering_comp_graph}
\end{figure}

Figure~\ref{fig: gradual_clustering_comp_graph} depicts the infection trends in each community as the coupling coefficient magnitude ($\epsilon$) increases. Several observations emerge, and we will explore and eliminate any other contributing parameters in subsequent sections of our study. In Fig.~\ref{fig: gradual_clustering_comp_graph} (A-C), we observe that communities exhibit a tendency to draw closer together.
The emergence of the second peak can be explained by the varying dynamics between communities with and without test kits. Non-test kit communities exhibit a higher susceptibility to infection, which results in a rapid ascent of the first peak, compelling an increased production of test kits. In response, the abundant availability of test kits efficiently suppresses the initial peak within the test kit-equipped communities.

However, as the first peak recedes within non-test kit communities, a critical issue unfolds - the scarcity of available test kits. This dearth of resources leads to the resurgence of infections within the test kit communities, culminating in the appearance of the second peak. This second peak is not observed in non-test kit communities as the number of susceptibles remaining within the population is much less after the big first peak. Notably, as the coupling coefficient ($\epsilon$) is increased,  communities equipped with test kits also converge, the second peak diminishes, and only the first peak remains, reflecting the intricate interplay between community interactions and the availability of test kits.

In Figure~\ref{fig: gradual_clustering_comp_graph}(D), we see that the second peak for the communities with test kits has almost completely vanished. The infection trends for communities without test kits are now closer. Figure~\ref{fig: gradual_clustering_comp_graph}(E) is when we can observe a synchronisation of communities with and without test kits amongst themselves. In Fig. ~\ref{fig: gradual_clustering_comp_graph}(F-H), it can be observed that the two synchronised infection trends start to move towards each other until finally, in Fig. ~\ref{fig: gradual_clustering_comp_graph}(I), the two trends also synchronise to form a global cluster.

\subsection{Studying \texorpdfstring{$E$ vs $\epsilon$ plots and corresponding clustering of $I_{\text{max}}$}{Studying E vs epsilon plots and corresponding clustering of Imax}}

\begin{figure}[htp!]
    \centerline{\includegraphics[width=1\columnwidth]{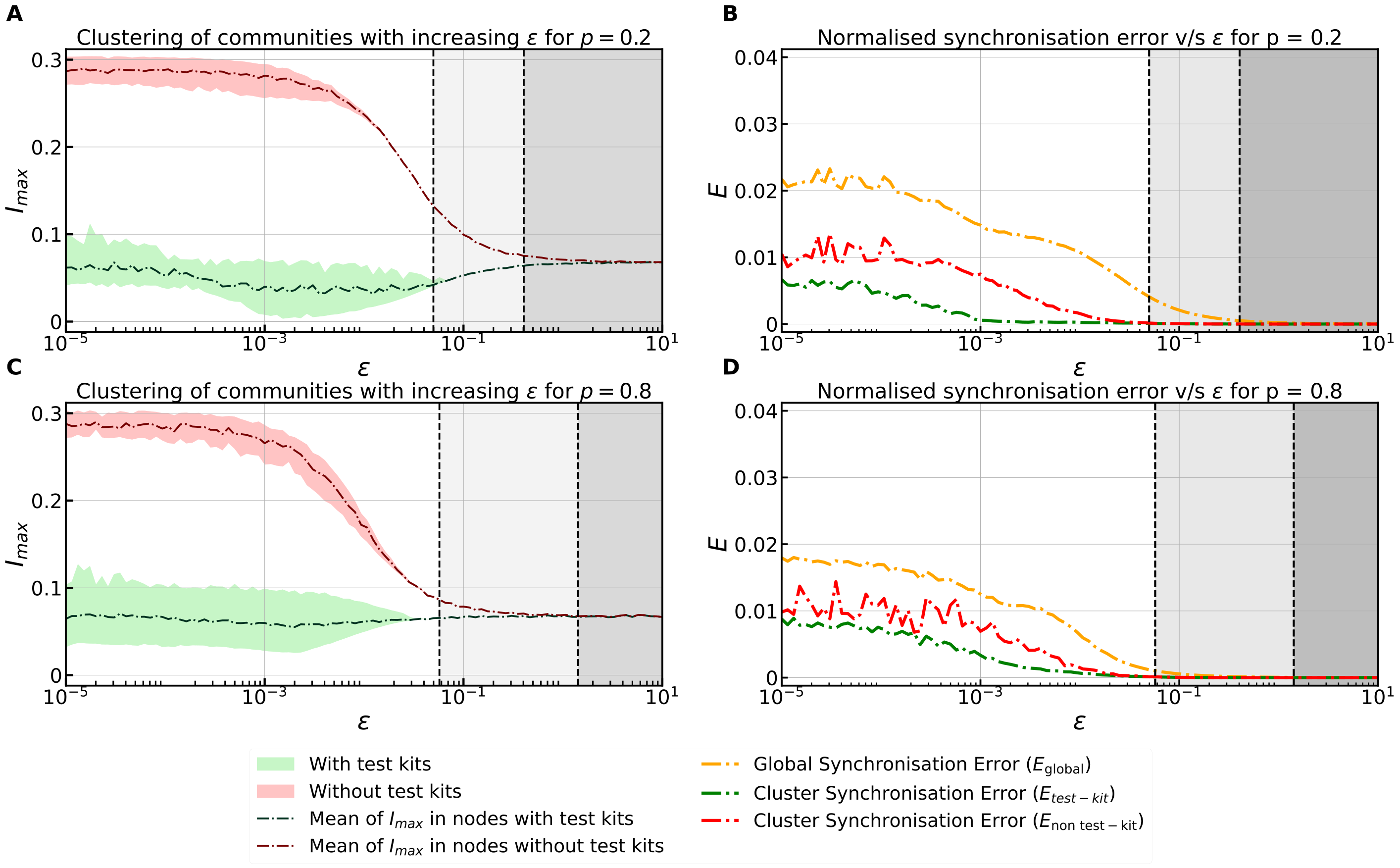}}
    \caption{Plots \textbf{A} and \textbf{C} show the spread of $I$\textsubscript{max} values for different communities decreasing with increasing $\epsilon$. Plots \textbf{B} and \textbf{D} show the decrease in Synchronisation Error with increasing $\epsilon$. There are 100 communities each with a population of $10000$ people. The plot parameters used were $\sigma = 0.1,\ \alpha_0 = 0.01\, \alpha_1 = 0.0001,\ \gamma = 0.07,\ \zeta = 0.02,\ \chi = 0.1$ and $\beta = 0.03$. $p$ denotes the fraction of communities having test kits. We can observe the existence of a bicluster in Plots \textbf{A} and \textbf{C} and confirm that the GSE vanishes only after CSE for both test-kit and testkit-free communities vanishes in plots \textbf{B} and \textbf{D}.}
    \label{fig: rmse_imax_epsilon}
\end{figure}

In Fig.~\ref{fig: rmse_imax_epsilon}(A, C) we report that $I$\textsubscript{max} for each community vs $\epsilon$ as the background patch, and the mean of $I$\textsubscript{max} across all communities is indicated by the solid lines through the patch. In Fig.~\ref{fig: rmse_imax_epsilon}(B, D), it is discernible that the $\epsilon$ values associated with errors $E$ falling below a pre-defined empirical threshold ($1.5 \times 10^{-4}$) partition the graph into three distinct regions. These regions are identified by colour in Fig.~\ref{fig: rmse_imax_epsilon}(B, D): the Multi-cluster region(characterized by a white backdrop), the Bi-cluster region(characterized by a light grey backdrop) when both test and non-testkit cluster errors are below the threshold but not the global error and finally, the Mono-cluster region(distinguished by a dark grey backdrop) where the global error is below the threshold as well. 
With increasing $\epsilon$, errors for test-kit and without test-kit node clusters vanish after a threshold epsilon while the Global Synchronisation Error is still non-zero.  Upon further increasing $\epsilon$, we observe complete clustering, and the Global Synchronisation Error (GSE) also drops to zero. We can see now that the GSE vanishes only after CSE for both test-kit and testkit-free communities vanishes establishing the existence of a bicluster region.

Figure~\ref{fig: rmse_imax_epsilon}(A, C) show that for smaller orders of the coupling coefficient $\epsilon$, the $I$\textsubscript{max} has a greater spread around its mean value - indicated by a patch around the line showing the mean value of $I$\textsubscript{max} across all communities. We can now see that the peak of infection curves form two clusters and then merge into one. Employing our formulation for synchronization error, we will ascertain that synchronization extends beyond the $I$\textsubscript{max} values alone, encompassing both the spread and the overall shape of the curve. Upon looking more closely, we observe that the mean $I$\textsubscript{max} for the non-testkit communities falls like a smooth-step function with increasing $\epsilon$, but the testkit communities first monotonically decrease and then after reaching a minimum start monotonically increasing, creating a small-dip in the plot. We can explain this using Fig.~\ref{fig: gradual_clustering_comp_graph}, where we observe two peaks for lower $\epsilon$ values in the test-kit communities, with the first peak being lower. As $\epsilon$ values increase, we observe the second peak reduces and the first peak increases in magnitude, leading to a particular epsilon value where the $I$\textsubscript{max} switches from the second to the first peak. As the first peak increases in magnitude and the second one decreases, $I$\textsubscript{max} initially decreases until the first peak overtakes the second peak to become the determinant of $I$\textsubscript{max} and continues increasing. Thus, we observe a dip at this point of overtake.

From this point on, we investigate the impact of other parameters in our study and confirm that the observed behaviour is due to $\epsilon$ alone. 





\subsection{Investigating Impact of the fraction of communities with test kits and reproductive number on clustering}
Let's take the fraction of communities with test kits to be denoted by $p$.
 We report the Synchronisation errors(Global and Cluster-wise) as a function of $\mathcal{R}_0$ and $\epsilon$ in Fig.4(A-C). The role of $p$ and $\epsilon$ is shown in Fig.4(D-F). In our analysis, $\mathcal{R}_0$ represents the basic reproduction number, calculated as the ratio of the transmission rate ($\beta$) and the constant $\alpha_0$.
In both Figure~\ref{fig: rmse_beta_epsilon_true_contourf}(A-C) and (D-F), our observations reveal a consistent trend wherein the Cluster-wise Synchronisation Error (CSE) converges to zero before the Global Synchronisation Error (GSE). This notable pattern persists across various values of $\mathcal{R}_0$ and $p$, indicating a prevalent emergence of the bicluster regime preceding the monocluster regime. Consequently, we confidently assert that the manifestation of the bicluster regime is solely attributed to the influence of the migration strength parameter, $\epsilon$.

\begin{figure}[htp!]
    \centerline{\includegraphics[width=1.1\columnwidth]{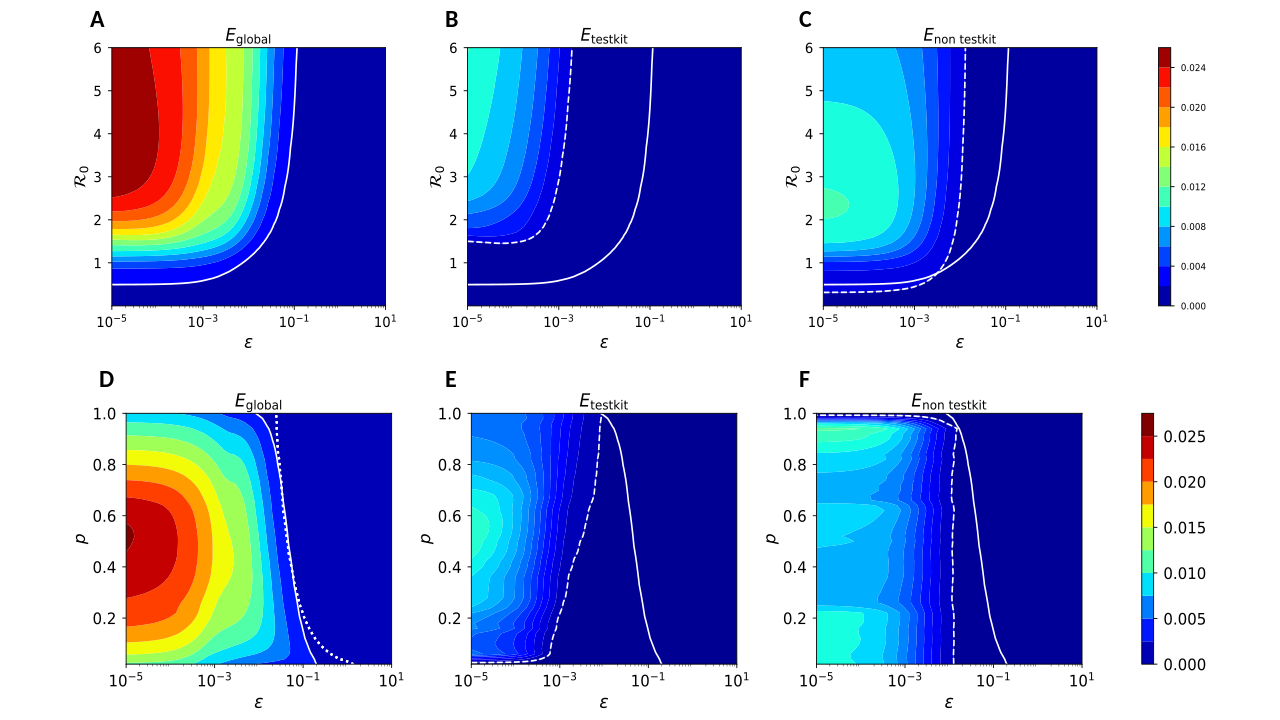}}
    \caption{Here we take 51 communities, each with a population of $10000$ people. The plot parameters used were $\sigma = 0.1,\ \alpha_0 = 0.01,\ \alpha_1 = 0.0001,\ \gamma = 0.07,\ \zeta = 0.02,\ \chi = 0.1$ and $\beta = 0.03$. The 2D projection of Synchronisation errors (Global and Cluster-wise) in the $\mathcal{R}_0$ vs $\epsilon$ plane are shown from (A-C) and in the $p$ vs $\epsilon$ plane are shown from (D-F). The dashed line indicates the contour line when the CSE vanishes, and the solid line indicates the same for GSE. Additionally, the dotted line in plot \textbf{(D)} shows the same threshold obtained via the reduced equation model.}
    \label{fig: rmse_beta_epsilon_true_contourf}
\end{figure}



\section{Reduced Equation model}
\label{sec: 4}
The observations suggest that test kit and non-test kit communities exhibit similar trends beyond a specific threshold value of $\epsilon$. Both \textit{test-kit} and \textit{non-test-kit} communities cluster, reducing the number of differential equations from $5M+1$ to only $11$. This simplification arises from consolidating the $M$ communities into just two, as beyond a certain $\epsilon$ value, distinct trends converge to either \textit{test-kit} or \textit{non-test-kit}.

\subsection{Derivation of reduced model equations}
To do this, we consolidate all the variables with equations containing a migration term (refer section 2.1) into a vector $\textbf{X}_{i}$ corresponding to the $i^{th}$ community. Let $y$ and $z$ be the number of communities with and without test kits, respectively.
$$ 
\textbf{X}_{i} = \begin{bmatrix}
  S_{i}\\
  E_{i}\\
  I_{i}\\
  R_{i}\\
\end{bmatrix} 
$$

We will start by looking at all $i \in [1, y]$, that is, all the communities where test kits were distributed. According to the equations in section 2.1, we can write the derivative of $\textbf{X}_i$ as

\begin{align}
\dot{\textbf{X}}_i &= f(\textbf{X}_i) + \frac{\epsilon}{N - 1}\displaystyle\sum_{j=1}^{N}{(\textbf{X}_j - \textbf{X}_i)}
\\ 
&= f(\textbf{X}_i) + \frac{\epsilon}{N - 1}{\biggl((\textbf{X}_1 - \textbf{X}_i) + (\textbf{X}_2 - \textbf{X}_i) + (\textbf{X}_3 - \textbf{X}_i) + \dots + (\textbf{X}_N - \textbf{X}_i)\biggr)}
\end{align}

Here, $f(\textbf{X}_i)$ denotes the non-migration terms for each equations for compartments present in $X_i$
$$
    f(\textbf{X}_i) = \begin{bmatrix}
        -\beta\left(\frac{S_iI_i}{N_i}\right)\\
        \beta\left(\frac{S_iI_i}{N_i}\right)- \sigma E_i\\
        \sigma E_i - \left(\alpha_0 + g_i(K)\right)I_i\\
        \gamma H_i
    \end{bmatrix}
$$

After entering the Bi-Cluster regime, communities corresponding to $i \in [1, y]$ and communities corresponding to $j \in [y + 1, N]$ must have clustered amongst themselves. This implies that for all $i\in [1, y]$ 
\begin{align}
    \textbf{X}_1 = \textbf{X}_2 = \textbf{X}_3 = \dots = \textbf{X}_y \implies \textbf{X}_i = \textbf{X}_y
\end{align}

Similarly for all $j\in [y+1, N]$,
\begin{align}
    \textbf{X}_{y+1} = \textbf{X}_{y+2} = \dots = \textbf{X}_{N} \implies \textbf{X}_j = \textbf{Y}_z
\end{align}

Many terms from the summation collapse together using the above equalities,

\begin{align}
    \dot{\textbf{X}}_y &= f(\textbf{X}_y) + \frac{\epsilon}{N - 1}\biggl({\displaystyle\sum_{j=1}^{y}{(\textbf{X}_j - \textbf{X}_y)}} + \displaystyle\sum_{j=y+1}^{N}{(\textbf{X}_j - \textbf{X}_y)}\biggr)\\
    &= f(\textbf{X}_y) + \frac{\epsilon}{N - 1}\biggl(\displaystyle\sum_{j=y+1}^{N}{(\textbf{X}_j - \textbf{X}_y)}\biggr)
\end{align}

Since $y + z = N$,

\begin{align}
    \dot{\textbf{X}}_y &= f(\textbf{X}_y) + \frac{\epsilon}{N - 1}\biggl ({\biggl(\displaystyle\sum_{j=y+1}^{N}{\textbf{X}_j}\biggr)} - z\times \textbf{X}_y \biggr)\\
    &= f(\textbf{X}_y) + \frac{\epsilon}{N - 1}\biggl({\biggl(\displaystyle\sum_{j=y+1}^{N}{\textbf{Y}_z}\biggr)} - z\times \textbf{X}_y \biggr)\\
    &= f(\textbf{X}_y) + \frac{\epsilon}{N - 1}\biggl(z\times \textbf{Y}_z - z\times \textbf{X}_y \biggr)\\
    &= f(\textbf{X}_y) + \frac{z\epsilon}{N - 1}\biggl(\textbf{Y}_z - \textbf{X}_y \biggr)
\end{align}

Without loss of generality, we can deduce a similar equation for non-test kit communities too,

\begin{align}
    &\dot{\textbf{Y}}_z = f(\textbf{Y}_z) + \frac{y\epsilon}{N - 1}\biggl(\textbf{X}_y - \textbf{Y}_z \biggr)
\end{align}

Along with these 8 equations, we have $2$ for $H_{i}$ and $H_{j}$ and $1$ more for $K$,
\begin{align}
    &\dot{H}_y = \alpha I_y + \gamma H_y \\
    &\dot{H}_z = \alpha I_z + \gamma H_z
\end{align}

Here, $H_y$ refers to the Hospitalization compartment for test kit communities and $H_z$ refers to the Hospitalization compartment for non-test kit communities. Similarly for $K$,
\begin{align}
    &\dot{K} = \zeta( yI_y + zI_z) - \chi K
\end{align}
Thus, we reduce $5 M + 1$ differential equations to 11 differential equations.

On plotting the infection values obtained from with and without test-kit equations, they follow very closely after the threshold $\epsilon$ when clustering begins as we expected.
\begin{figure}[htp!]
    \centerline{\includegraphics[width=1\columnwidth]{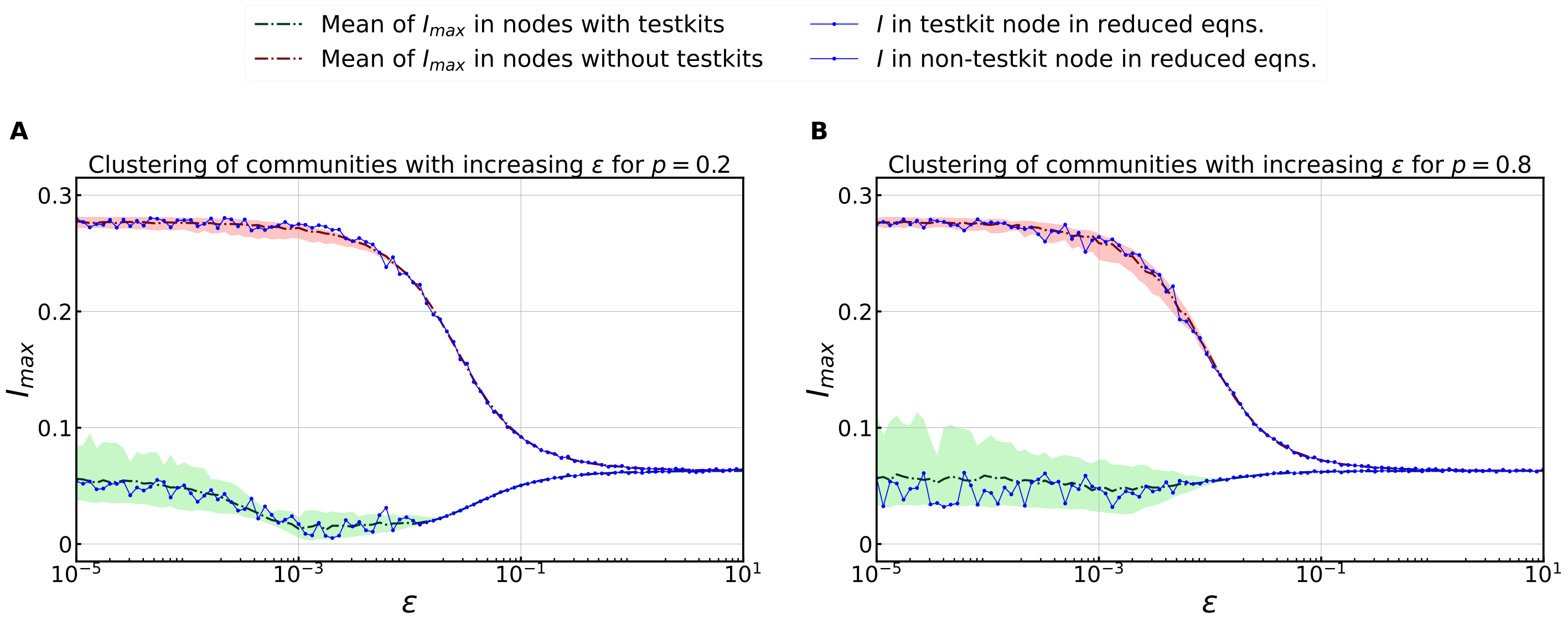}}
    \caption{Plotting $I$\textsubscript{max} obtained from simulation and reduced equation model together to see areas of synchronisation. There are $100$ communities with a population of $10,000$ people each. 0.2 and 0.8 fractions of communities get test kits. Initial conditions were randomised as a uniform distribution between 1 and 1000. The plot parameters used were $\sigma = 0.1,\ \alpha_0 = 0.01,\ \alpha_1 = 0.0001,\ \gamma = 0.07,\ \zeta = 0.02,\ \chi = 0.1$ and $\beta = 0.03$. }
    \label{fig: complete_red_model}
\end{figure}

 Figure~\ref{fig: complete_red_model} shows that the reduced model poorly approximates the value of $\epsilon$ before entering the Bi-cluster regime. Part of the possible reasons behind this poor approximation at low coupling strength comes from the assumed synchronisation before the clusters actually form. However, the trend of the reduced equation model closely follows the real model at higher values of $\epsilon$ due to the formation of clusters.

\section{Impact of the coupling coefficient on an Erdos-Renyi Network}
\label{sec: 5}


In this investigation, we experimentally validate our conceptual framework on an Erdos-Renyi Network to ascertain the presence of a bicluster regime. Consistent with the observations in the preceding section, the clustering dynamics in this network exhibit the formation of bi-clusters, culminating in complete synchronisation. A nuanced distinction arises when compared to the fully connected network; specifically, the two clusters within the formed bicluster display subtle noise. This variation introduces an additional layer of complexity in the clustering behaviour, highlighting the network topology's influence on the observed synchronisation patterns.

\begin{figure}[htp!]
    \centerline{\includegraphics[width=1\columnwidth]{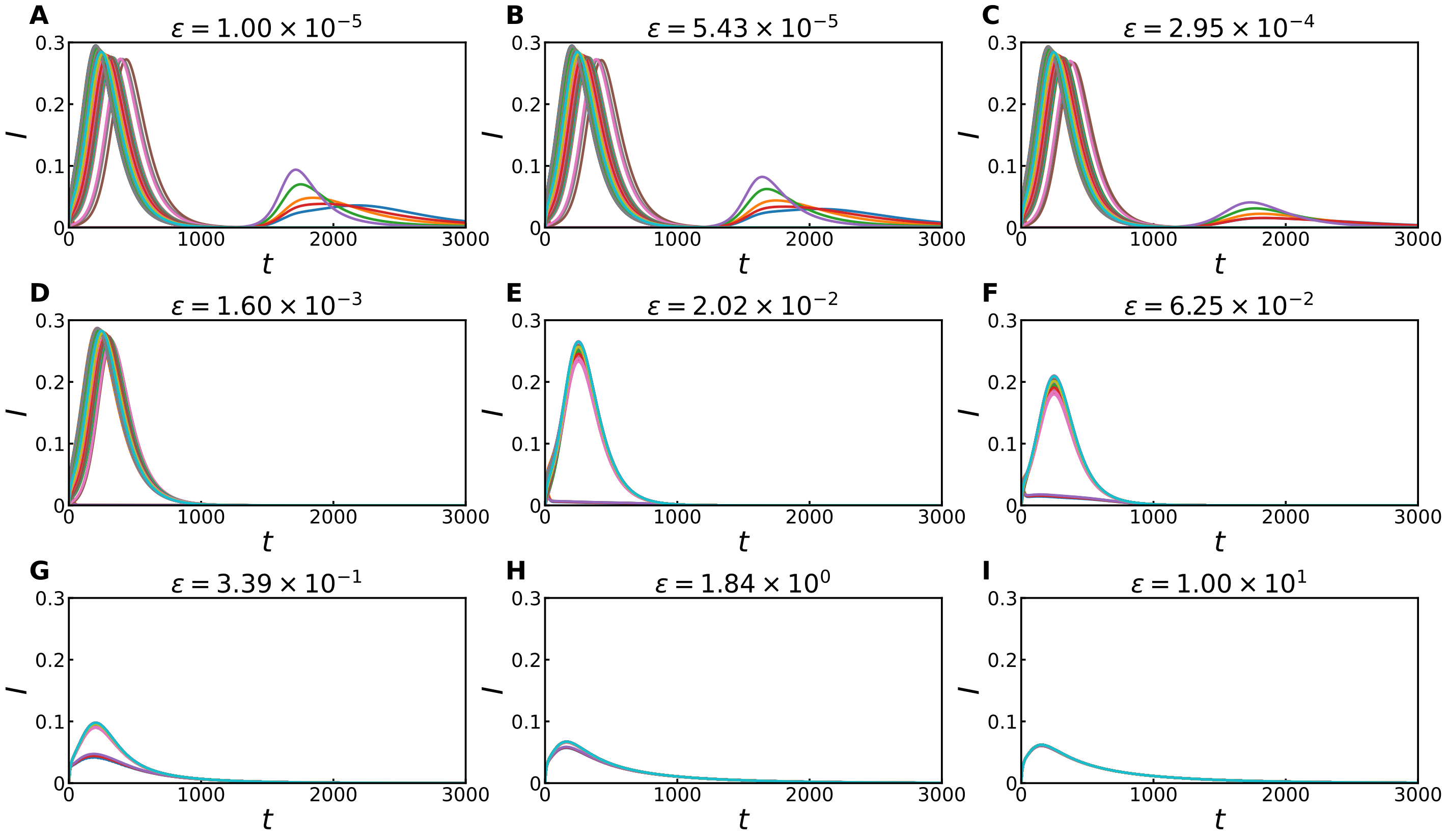}}
    \caption{There are $100$ communities with a population of $10000$ people, each. Test kits were distributed among $40\%$ of the communities. Initial conditions were randomised as uniform distribution between 1 and 1000. The plot parameters used were $\sigma = 0.1,\ \alpha_0 = 0.01,\ \alpha_1 = 0.0001,\ \gamma = 0.07,\ \zeta = 0.02,\ \chi = 0.1$ and $\beta = 0.03$. We observe a synchronisation of infection trends with an increasing coupling coefficient. This aligns with what we see in Figure ~\ref{fig: gradual_clustering_comp_graph}, where clusters show some noise near the peaks. Despite this, a clear synchronization pattern remains evident.}
    \label{fig: gradual_clustering_er}
\end{figure}

\subsection{Derivation of reduced model equations}

In alignment with the derivation for a complete graph, we pursue a comparable reduced model formulation. This requires the assumption that the degree in each node is approximately equal to the average degree across all nodes, denoted as $\langle d\rangle$.To do this, we again consolidate all the variables with equations containing a migration term (refer section 2.1) into a vector $\textbf{X}_{i}$ corresponding to the $i^{th}$ community. Let $y$ and $z$ be the number of communities with and without test kits, respectively.
$$ 
\textbf{X}_{i} = \begin{bmatrix}
  S_{i}\\
  E_{i}\\
  I_{i}\\
  R_{i}\\
\end{bmatrix} 
$$

Our derivation starts by looking at all $i \in [1, y]$, that is, all the communities where test kits were distributed. According to the equations in section 2.1, we can write the derivative of $\textbf{X}_i$ as

\begin{align}
\dot{\textbf{X}}_i &= f(\textbf{X}_i) + \frac{\epsilon}{N - 1}\displaystyle\sum_{j=1}^{N}{(\textbf{X}_j - \textbf{X}_i)}
\\
&= f(\textbf{X}_i) + \frac{\epsilon}{d_i}{\biggl((\textbf{X}_1 - \textbf{X}_i) + (\textbf{X}_2 - \textbf{X}_i) + (\textbf{X}_3 - \textbf{X}_i) + \dots + (\textbf{X}_N - \textbf{X}_i)\biggr)}
\end{align}

\begin{align}
\dot{\textbf{X}}_i 
&\approx f(\textbf{X}_i) + \frac{\epsilon}{\langle d\rangle}{\biggl((\textbf{X}_1 - \textbf{X}_i) + (\textbf{X}_2 - \textbf{X}_i) + (\textbf{X}_3 - X_i) + \dots + (\textbf{X}_N - \textbf{X}_i)\biggr)}
\end{align}


The main difference lies in how we approximate the degree of each node in the network to its expected value, denoted as $\langle d\rangle$. This approximation is based on the Poissonian model applied to the Erdos-Renyi network. This choice reflects our aim to capture the network's inherent structure, emphasising the utility of the Poissonian model in representing the expected degree distribution.

\begin{figure}[htp!]    
    \centerline{\includegraphics[width=1\columnwidth]{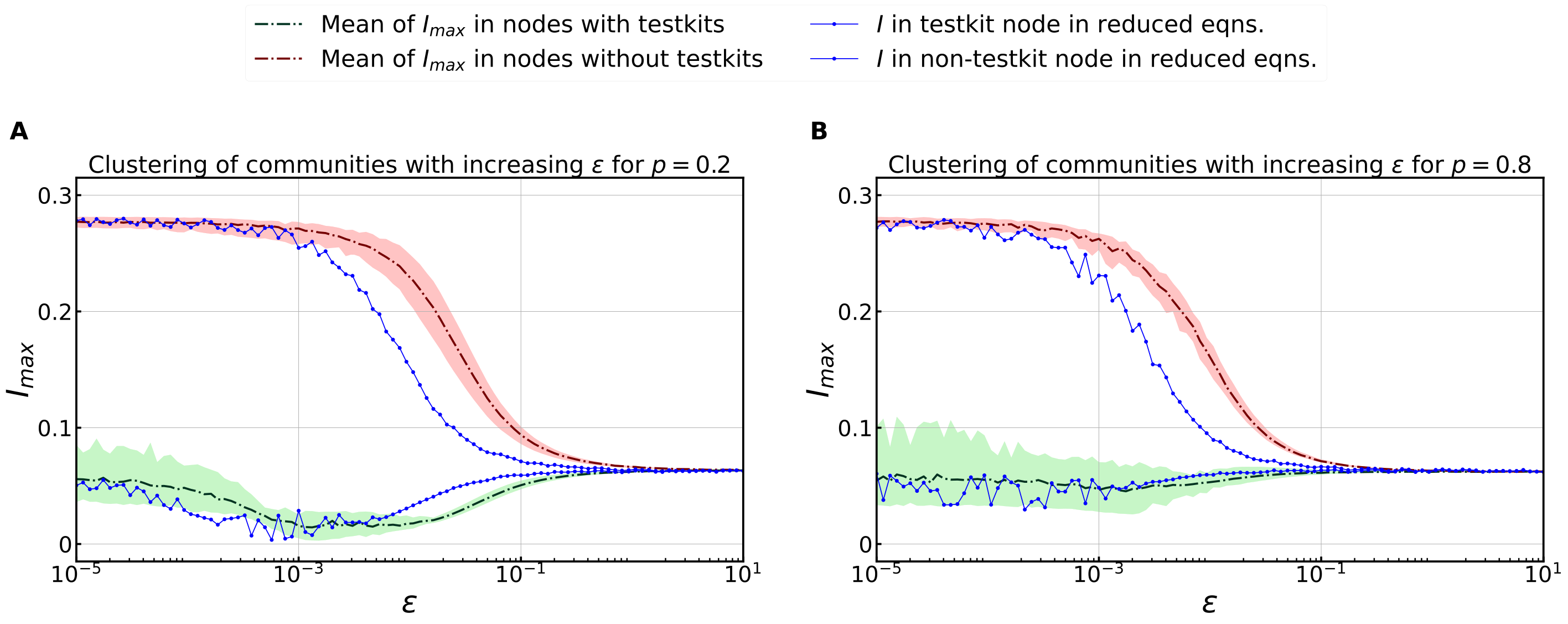}}
    \caption{Plotting $I$\textsubscript{max} obtained from simulation and reduced equation model together to see areas of synchronisation. There are $100$ communities with a population of $10,000$ people each. 0.2 and 0.8 fractions of communities get test kits. The plots here are for an Erdos-Renyi network. Initial conditions were randomised as a uniform distribution between 1 and 1000. The plot parameters used were $\sigma = 0.1,\ \alpha_0 = 0.01,\ \alpha_1 = 0.0001,\ \gamma = 0.07,\ \zeta = 0.02,\ \chi = 0.1$ and $\beta = 0.03$. }
    \label{fig: red_model_er}
\end{figure}

Similar to our earlier derivation for the complete graph, we can condense all differential equations into the previously discussed 11 equations. This simplification provides a more concise representation of the system dynamics, following the approach employed in our analysis of the complete graph.

\begin{align}
    &\dot{\textbf{X}}_y = f(\textbf{X}_y) + \frac{z\epsilon}{\langle d\rangle}\biggl(\textbf{Y}_z - \textbf{X}_y \biggr) \\
    &\dot{\textbf{Y}}_z = f(\textbf{Y}_z) + \frac{y\epsilon}{\langle d\rangle}\biggl(\textbf{X}_y - \textbf{Y}_z \biggr) \\
    &\dot{H}_y = \alpha I_y + \gamma H_y \\
    &\dot{H}_z = \alpha I_z + \gamma H_z \\
    &\dot{K} = \zeta( yI_y + zI_z) - \chi K
\end{align}

\subsection{Comparing infection trends with reduced model prediction}

The deviation of the behaviour of the reduced model in the bi-cluster regime becomes increasingly apparent upon inspecting the plots. The reduced model lags behind the true behaviour of the two clusters. As shown in the derivation of the reduced model equation, this follows from approximating the degree distribution to be exactly equal to its expected value.

\begin{figure}[htp!]
    \centerline{\includegraphics[width=1\columnwidth]{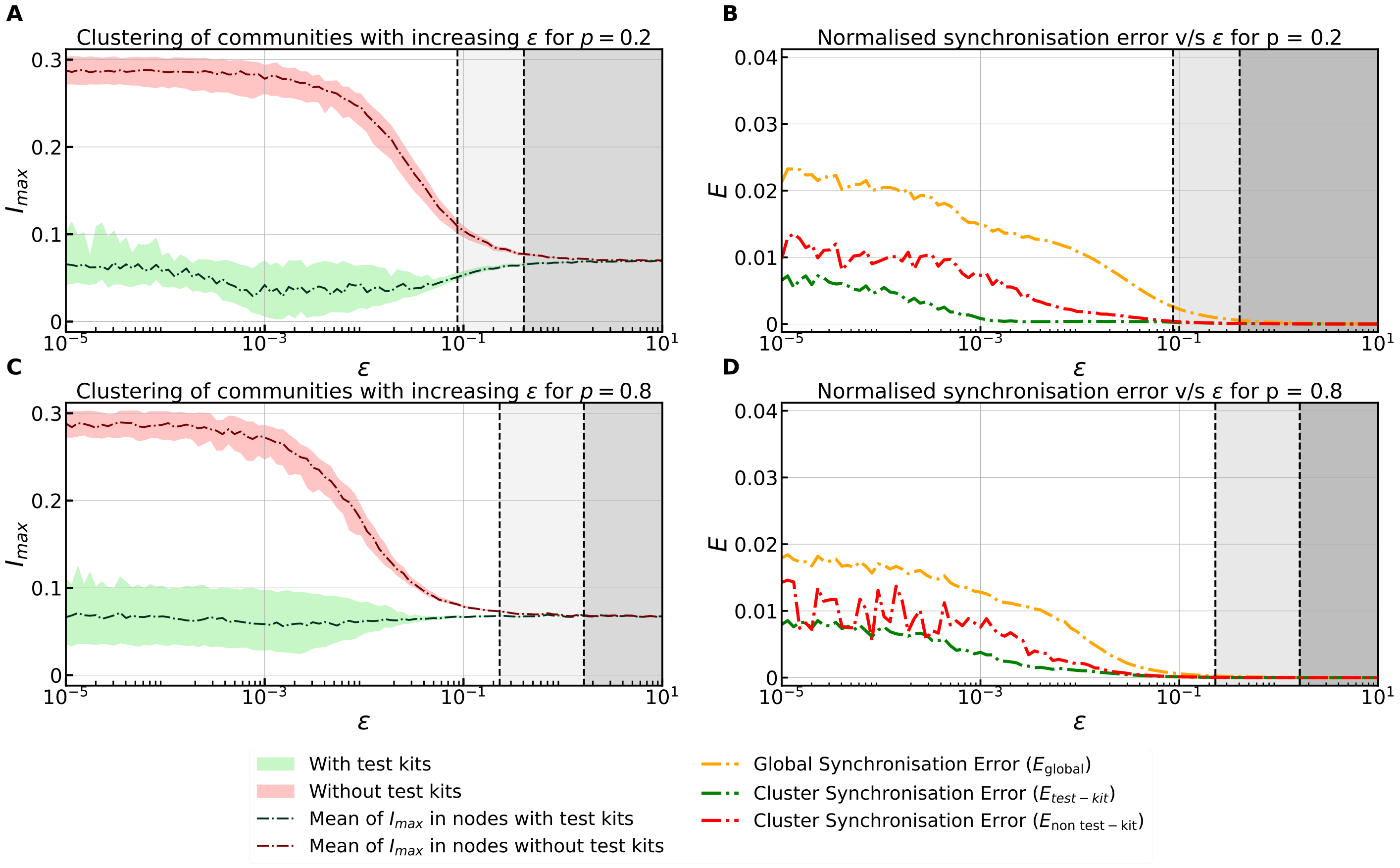}}
    \caption{There are $51$ communities with a population of $10000$ people, each. 0.4 fraction of communities got test kits. Initial conditions were randomised as a uniform distribution between 1 and 1000. This bears resemblance with what was observed in Fig.~\ref{fig: rmse_imax_epsilon}.}
    \label{fig: rmse_imax_epsilon_er}
\end{figure}

As previously observed, the synchronisation error for individual clusters diminishes before the Global Synchronisation Error reaches zero, aligning with our anticipated pattern. Consequently, we assert the definitive presence of a bi-cluster phenomenon in the Erdos-Renyi graph. In Figure~\ref{fig: rmse_imax_epsilon_er}, the bi-cluster regime manifests over a more confined range of $\epsilon$ than noted in the case of a complete graph.


\section{Conclusion}
\label{sec: 6}
The efficacy of the test-kit intervention strategy is evident, particularly when deploying test-kits in nodes with high degrees. Our study demonstrates that this approach, coupled with a substantial coupling strength, effectively mitigates and synchronises infection trends within the population. We reveal a critical range of coupling coefficients wherein infection trends bifurcate into two distinct clusters based on the presence or absence of test-kit distribution.

Furthermore, our investigation establishes that synchronisation is solely contingent on coupling strength. We analyse this synchronisation across various fractions of distributed test-kits and different transmission rates, emphasising its dependency on coupling strength. The observed synchronisation behaviour facilitates the reduction of numerous differential equations to a mere 11, offering a simplified model applicable to both complete graphs and Erdos-Renyi formulations. This approximation proves effective, especially as the coupling coefficient enters the bi-cluster regime. Reducing the number of differential equations streamlines the model and enhances its practicality in the face of increasing complexity. 

Future directions of work could involve carrying out the outlined analysis for scale-free networks. Additionally, delving into the captivating landscape of multilayer graphs, where nodes hold dual opinions, promises to offer a richer understanding of how infections propagate across diverse layers of influence. These intriguing directions not only expand the breadth of our research but also hold the potential to unveil novel insights into the complex dynamics of contagion within evolving network landscapes.




\printbibliography

@article{anderson2004epidemiology,
  title={Epidemiology, transmission dynamics and control of SARS: the 2002--2003 epidemic},
  author={Anderson, Roy M and Fraser, Christophe and Ghani, Azra C and Donnelly, Christl A and Riley, Steven and Ferguson, Neil M and Leung, Gabriel M and Lam, Tai H and Hedley, Anthony J},
  journal={Philosophical Transactions of the Royal Society of London. Series B: Biological Sciences},
  volume={359},
  number={1447},
  pages={1091--1105},
  year={2004},
  publisher={The Royal Society}
}

@article{belik2011natural,
  title={Natural human mobility patterns and spatial spread of infectious diseases},
  author={Belik, Vitaly and Geisel, Theo and Brockmann, Dirk},
  journal={Physical Review X},
  volume={1},
  number={1},
  pages={011001},
  year={2011},
  publisher={APS}
}

@article{khanra2022endowing,
  title = {Endowing networks with desired symmetries and modular behavior},
  author = {Khanra, P. and Ghosh, S. and Aleja, D. and Alfaro-Bittner, K. and Contreras-Aso, G. and Criado, R. and Romance, M. and Boccaletti, S. and Pal, P. and Hens, C.},
  journal = {Phys. Rev. E},
  volume = {108},
  issue = {5},
  pages = {054309},
  numpages = {7},
  year = {2023},
  publisher = {American Physical Society},
  doi = {10.1103/PhysRevE.108.054309},
  url = {https://link.aps.org/doi/10.1103/PhysRevE.108.054309}
}

@article{rakshit2023predicting,
  title={Predicting aging transition using Echo state network.},
  author={Rakshit, B and Kartha, AJ and Hens, C and others},
  journal={Chaos (Woodbury, NY)},
  volume={33},
  number={8},
  pages={081102--081102},
  year={2023}
}

@article{hens2015bursting,
  title={Bursting dynamics in a population of oscillatory and excitable Josephson junctions},
  author={Hens, Chittaranjan and Pal, Pinaki and Dana, Syamal K},
  journal={Physical Review E},
  volume={92},
  number={2},
  pages={022915},
  year={2015},
  publisher={APS}
}

@article{saha2predicting,
  title={Predicting bursting in a complete graph of mixed population through reservoir computing},
  author={Saha, Suman and Mishra, Arindam and Ghosh, Subrata and Dana, Syamal K and Hens, Chittaranjan},
  journal={Phys Rev Research},
  volume={2},
  pages={033338},
  publisher={American Physical Society}
}

@article{saha2020infection,
  title={Infection spreading and recovery in a square lattice.},
  author={Saha, S and Mishra, A and Dana, SK and Hens, C and Bairagi, N},
  journal={Physical review. E},
  volume={102},
  number={5-1},
  pages={052307--052307},
  year={2020}
}

@article{jalan2016cluster,
  title={Cluster synchronization in multiplex networks},
  author={Jalan, Sarika and Singh, Aradhana},
  journal={Europhysics Letters},
  volume={113},
  number={3},
  pages={30002},
  year={2016},
  publisher={IOP Publishing}
}

@article{sorrentino2016complete,
  title={Complete characterization of the stability of cluster synchronization in complex dynamical networks},
  author={Sorrentino, Francesco and Pecora, Louis M and Hagerstrom, Aaron M and Murphy, Thomas E and Roy, Rajarshi},
  journal={Science advances},
  volume={2},
  number={4},
  pages={e1501737},
  year={2016},
  publisher={American Association for the Advancement of Science}
}

@article{pecora2014cluster,
  title={Cluster synchronization and isolated desynchronization in complex networks with symmetries},
  author={Pecora, Louis M and Sorrentino, Francesco and Hagerstrom, Aaron M and Murphy, Thomas E and Roy, Rajarshi},
  journal={Nature communications},
  volume={5},
  number={1},
  pages={4079},
  year={2014},
  publisher={Nature Publishing Group UK London}
}

@article{mishra2023chimeras,
  title={Chimeras in globally coupled oscillators: A review},
  author={Mishra, Arindam and Saha, Suman and Dana, Syamal K},
  journal={Chaos: An Interdisciplinary Journal of Nonlinear Science},
  volume={33},
  number={9},
  year={2023},
  publisher={AIP Publishing}
}

@article{coburn2009modeling,
  title={Modeling influenza epidemics and pandemics: insights into the future of swine flu (H1N1)},
  author={Coburn, Brian J and Wagner, Bradley G and Blower, Sally},
  journal={BMC medicine},
  volume={7},
  number={1},
  pages={1--8},
  year={2009},
  publisher={BioMed Central}
}

@article{dye2003modeling,
  title={Modeling the SARS epidemic},
  author={Dye, Chris and Gay, Nigel},
  journal={Science},
  volume={300},
  number={5627},
  pages={1884--1885},
  year={2003},
  publisher={American Association for the Advancement of Science}
}

@article{khanra2022identifying,
  title={Identifying symmetries and predicting cluster synchronization in complex networks},
  author={Khanra, Pitambar and Ghosh, Subrata and Alfaro-Bittner, Karin and Kundu, Prosenjit and Boccaletti, Stefano and Hens, Chittaranjan and Pal, Pinaki},
  journal={Chaos, Solitons \& Fractals},
  volume={155},
  pages={111703},
  year={2022},
  publisher={Elsevier}
}

@article{ghosh2020emergence,
  title={Emergence of mixed mode oscillations in random networks of diverse excitable neurons: the role of neighbors and electrical coupling},
  author={Ghosh, Subrata and Mondal, Argha and Ji, Peng and Mishra, Arindam and Dana, Syamal K and Antonopoulos, Chris G and Hens, Chittaranjan},
  journal={Frontiers in Computational Neuroscience},
  volume={14},
  pages={49},
  year={2020},
  publisher={Frontiers Media SA}
}

@article{kundu2017transition,
  title={Transition to synchrony in degree-frequency correlated Sakaguchi-Kuramoto model},
  author={Kundu, Prosenjit and Khanra, Pitambar and Hens, Chittaranjan and Pal, Pinaki},
  journal={Physical Review E},
  volume={96},
  number={5},
  pages={052216},
  year={2017},
  publisher={APS}
}

@article{khanra2018explosive,
  title={Explosive synchronization in phase-frustrated multiplex networks},
  author={Khanra, Pitambar and Kundu, Prosenjit and Hens, Chittaranjan and Pal, Pinaki},
  journal={Physical Review E},
  volume={98},
  number={5},
  pages={052315},
  year={2018},
  publisher={APS}
}

@article{lahav2022topological,
  title={Topological synchronization of chaotic systems},
  author={Lahav, Nir and Sendi{\~n}a-Nadal, Irene and Hens, Chittaranjan and Ksherim, Baruch and Barzel, Baruch and Cohen, Reuven and Boccaletti, Stefano},
  journal={Scientific reports},
  volume={12},
  number={1},
  pages={2508},
  year={2022},
  publisher={Nature Publishing Group UK London}
}

@article{kapitaniak2012synchronization,
  title={Synchronization of clocks},
  author={Kapitaniak, Marcin and Czolczynski, Krzysztof and Perlikowski, Przemys{\l}aw and Stefanski, Andrzej and Kapitaniak, Tomasz},
  journal={Physics Reports},
  volume={517},
  number={1-2},
  pages={1--69},
  year={2012},
  publisher={Elsevier}
}

@article{cote2017behavioural,
  title={Behavioural synchronization of large-scale animal movements--disperse alone, but migrate together?},
  author={Cote, Julien and Bocedi, Greta and Debeffe, Lucie and Chudzi{\'n}ska, Magda E and Weigang, Helene C and Dytham, Calvin and Gonzalez, Georges and Matthysen, Erik and Travis, Justin and Baguette, Michel and others},
  journal={Biological reviews},
  volume={92},
  number={3},
  pages={1275--1296},
  year={2017},
  publisher={Wiley Online Library}
}

@article{hopfield1995rapid,
  title={Rapid local synchronization of action potentials: toward computation with coupled integrate-and-fire neurons.},
  author={Hopfield, John J and Herz, Andreas V},
  journal={Proceedings of the National Academy of Sciences},
  volume={92},
  number={15},
  pages={6655--6662},
  year={1995},
  publisher={National Acad Sciences}
}

@article{ivanov2009maternal,
  title={Maternal--fetal heartbeat phase synchronization},
  author={Ivanov, Plamen Ch and Ma, Qianli DY and Bartsch, Ronny P},
  journal={Proceedings of the National Academy of Sciences},
  volume={106},
  number={33},
  pages={13641--13642},
  year={2009},
  publisher={National Acad Sciences}
}

@article{nagy2018synchronization,
  title={Synchronization, coordination and collective sensing during thermalling flight of freely migrating white storks},
  author={Nagy, M{\'a}t{\'e} and Couzin, Iain D and Fiedler, Wolfgang and Wikelski, Martin and Flack, Andrea},
  journal={Philosophical Transactions of the Royal Society B: Biological Sciences},
  volume={373},
  number={1746},
  pages={20170011},
  year={2018},
  publisher={The Royal Society}
}

@article{ramirez2019modeling,
  title={Modeling fireflies synchronization},
  author={Ram{\'\i}rez-{\'A}vila, Gonzalo Marcelo and Kurths, J{\"u}rgen and Depickere, St{\'e}phanie and Deneubourg, Jean-Louis},
  journal={A mathematical modeling approach from nonlinear dynamics to complex systems},
  pages={131--156},
  year={2019},
  publisher={Springer}
}

@article{boccaletti2002synchronization,
  title={The synchronization of chaotic systems},
  author={Boccaletti, Stefano and Kurths, J{\"u}rgen and Osipov, Grigory and Valladares, DL and Zhou, CS},
  journal={Physics reports},
  volume={366},
  number={1-2},
  pages={1--101},
  year={2002},
  publisher={Elsevier}
}

@article{arenas2008synchronization,
  title={Synchronization in complex networks},
  author={Arenas, Alex and D{\'\i}az-Guilera, Albert and Kurths, Jurgen and Moreno, Yamir and Zhou, Changsong},
  journal={Physics reports},
  volume={469},
  number={3},
  pages={93--153},
  year={2008},
  publisher={Elsevier}
}

@article{nielsen2021covid,
  title={COVID-19 superspreading suggests mitigation by social network modulation},
  author={Nielsen, Bjarke Frost and Simonsen, Lone and Sneppen, Kim},
  journal={Physical Review Letters},
  volume={126},
  number={11},
  pages={118301},
  year={2021},
  publisher={APS}
}

@article{james2007event,
  title={An event-based model of superspreading in epidemics},
  author={James, Alex and Pitchford, Jonathan W and Plank, Michael J},
  journal={Proceedings of the Royal Society B: Biological Sciences},
  volume={274},
  number={1610},
  pages={741--747},
  year={2007},
  publisher={The Royal Society London}
}

@article{hindes2022outbreak,
  title={Outbreak size distribution in stochastic epidemic models},
  author={Hindes, Jason and Assaf, Michael and Schwartz, Ira B},
  journal={Physical Review Letters},
  volume={128},
  number={7},
  pages={078301},
  year={2022},
  publisher={APS}
}

@article{hindes2016epidemic,
  title={Epidemic extinction and control in heterogeneous networks},
  author={Hindes, Jason and Schwartz, Ira B},
  journal={Physical review letters},
  volume={117},
  number={2},
  pages={028302},
  year={2016},
  publisher={APS}
}

@article{holme2013extinction,
  title={Extinction times of epidemic outbreaks in networks},
  author={Holme, Petter},
  journal={PloS one},
  volume={8},
  number={12},
  pages={e84429},
  year={2013},
  publisher={Public Library of Science San Francisco, USA}
}

@misc{wang2017vaccination,
  title={Vaccination and epidemics in networked populations—an introduction},
  author={Wang, Zhen and Moreno, Yamir and Boccaletti, Stefano and Perc, Matja{\v{z}}},
  journal={Chaos, Solitons \& Fractals},
  volume={103},
  pages={177--183},
  year={2017},
  publisher={Elsevier}
}

@article{wang2016statistical,
  title={Statistical physics of vaccination},
  author={Wang, Zhen and Bauch, Chris T and Bhattacharyya, Samit and d'Onofrio, Alberto and Manfredi, Piero and Perc, Matja{\v{z}} and Perra, Nicola and Salath{\'e}, Marcel and Zhao, Dawei},
  journal={Physics Reports},
  volume={664},
  pages={1--113},
  year={2016},
  publisher={Elsevier}
}

@article{hens2019spatiotemporal,
  title={Spatiotemporal signal propagation in complex networks},
  author={Hens, Chittaranjan and Harush, Uzi and Haber, Simi and Cohen, Reuven and Barzel, Baruch},
  journal={Nature Physics},
  volume={15},
  number={4},
  pages={403--412},
  year={2019},
  publisher={Nature Publishing Group UK London}
}

@article{iannelli2017effective,
  title={Effective distances for epidemics spreading on complex networks},
  author={Iannelli, Flavio and Koher, Andreas and Brockmann, Dirk and H{\"o}vel, Philipp and Sokolov, Igor M},
  journal={Physical Review E},
  volume={95},
  number={1},
  pages={012313},
  year={2017},
  publisher={APS}
}

@article{pastor2002epidemic,
  title={Epidemic dynamics in finite size scale-free networks},
  author={Pastor-Satorras, Romualdo and Vespignani, Alessandro},
  journal={Physical Review E},
  volume={65},
  number={3},
  pages={035108},
  year={2002},
  publisher={APS}
}

@article{pastor2015epidemic,
  title={Epidemic processes in complex networks},
  author={Pastor-Satorras, Romualdo and Castellano, Claudio and Van Mieghem, Piet and Vespignani, Alessandro},
  journal={Reviews of modern physics},
  volume={87},
  number={3},
  pages={925},
  year={2015},
  publisher={APS}
}

@article{pastor2001epidemic,
  title={Epidemic spreading in scale-free networks},
  author={Pastor-Satorras, Romualdo and Vespignani, Alessandro},
  journal={Physical review letters},
  volume={86},
  number={14},
  pages={3200},
  year={2001},
  publisher={APS}
}

@article{simonsen1998pandemic,
  title={Pandemic versus epidemic influenza mortality: a pattern of changing age distribution},
  author={Simonsen, Lone and Clarke, Matthew J and Schonberger, Lawrence B and Arden, Nancy H and Cox, Nancy J and Fukuda, Keiji},
  journal={Journal of infectious diseases},
  volume={178},
  number={1},
  pages={53--60},
  year={1998},
  publisher={The University of Chicago Press}
}

@article{akin2020understanding,
  title={Understanding dynamics of pandemics},
  author={Akin, Levent and G{\"o}zel, Mustafa G{\"o}khan},
  journal={Turkish journal of medical sciences},
  volume={50},
  number={9},
  pages={515--519},
  year={2020}
}

@article{preti2020psychological,
  title={The psychological impact of epidemic and pandemic outbreaks on healthcare workers: rapid review of the evidence},
  author={Preti, Emanuele and Di Mattei, Valentina and Perego, Gaia and Ferrari, Federica and Mazzetti, Martina and Taranto, Paola and Di Pierro, Rossella and Madeddu, Fabio and Calati, Raffaella},
  journal={Current psychiatry reports},
  volume={22},
  pages={1--22},
  year={2020},
  publisher={Springer}
}

@article{van2023attacks,
  title={Attacks on health care workers in historical pandemics and COVID-19},
  author={van Stekelenburg, Brett CA and De Cauwer, Harald and Barten, Dennis G and Mortelmans, Luc J},
  journal={Disaster medicine and public health preparedness},
  volume={17},
  pages={e309},
  year={2023},
  publisher={Cambridge University Press}
}

@article{brockmann2013hidden,
  title={The hidden geometry of complex, network-driven contagion phenomena},
  author={Brockmann, Dirk and Helbing, Dirk},
  journal={science},
  volume={342},
  number={6164},
  pages={1337--1342},
  year={2013},
  publisher={American Association for the Advancement of Science}
}

@article{maier2020effective,
  title={Effective containment explains subexponential growth in recent confirmed COVID-19 cases in China},
  author={Maier, Benjamin F and Brockmann, Dirk},
  journal={Science},
  volume={368},
  number={6492},
  pages={742--746},
  year={2020},
  publisher={American Association for the Advancement of Science}
}

@article{ndwandwe2021covid,
  title={COVID-19 vaccines},
  author={Ndwandwe, Duduzile and Wiysonge, Charles S},
  journal={Current opinion in immunology},
  volume={71},
  pages={111--116},
  year={2021},
  publisher={Elsevier}
}

@article{ciotti2020covid,
  title={The COVID-19 pandemic},
  author={Ciotti, Marco and Ciccozzi, Massimo and Terrinoni, Alessandro and Jiang, Wen-Can and Wang, Cheng-Bin and Bernardini, Sergio},
  journal={Critical reviews in clinical laboratory sciences},
  volume={57},
  number={6},
  pages={365--388},
  year={2020},
  publisher={Taylor \& Francis}
}

@article{jamieson2009h1n1,
  title={H1N1 2009 influenza virus infection during pregnancy in the USA},
  author={Jamieson, Denise J and Honein, Margaret A and Rasmussen, Sonja A and Williams, Jennifer L and Swerdlow, David L and Biggerstaff, Matthew S and Lindstrom, Stephen and Louie, Janice K and Christ, Cara M and Bohm, Susan R and others},
  journal={The Lancet},
  volume={374},
  number={9688},
  pages={451--458},
  year={2009},
  publisher={Elsevier}
}

@article{paper1,
    author={Subrata Ghosh and Abhishek Senapati and Joydev Chattopadhyay and Chittaranjan Hens and Dibakar Ghosh},
    title={Optimal test-kit-based intervention strategy of epidemic spreading in heterogeneous complex networks},
    year={2021},
    Journal={Chaos},    
}

@article{ paper3,
    author = {Jorge P. Rodríguez and  Víctor M. Eguíluz},
    title = {Coupling between infectious diseases leads to
synchronization of their dynamics},
    journal = {Chaos} ,
    year = {2023} 
}

@article{paper4,
    author = {Samali Ghosh and Arnob Ray and Dibakar Ghosh and Syamal K. Dana and
Tomasz Kapitaniak and Chittaranjan Hens and Ulrike Feudel} ,
    title = {Dispersal induced persistent dynamics in interacting ecological patches under
randomness of network topology},
    year = {2023}
}

@article{paper5,
    author = {Sanjeev Kumar Sharma and Argha Mondal and Arnab Mondal and Ranjit Kumar Upadhyay and Chittaranjan Hens},
    title = {Emergence of bursting in a network of memory dependent excitable and spiking leech-heart neurons},
    journal = {J. R. Soc. Interface},
    year = {2020}
}

@article{paper15,
    author = {Andrew J Tatem 1, Simon I Hay, David J Rogers},
    title = {Global traffic and disease vector dispersal},
    journal = {J. R. Soc. Interface},
    year = {2024}
}

@article{paper16,
    author = {John Reju Sam, Miller Joel C., Muylaert Renata L. and Hayman David T. S.},
    title = {High connectivity and human movement limits the impact of travel time on infectious disease transmission},
    journal = {Proceedings of the National Academy of Sciences of the United States of America},
    year = {2006}
}

@article{paper17,
    author = {Joseph T Wu 1, Kathy Leung 2, Gabriel M Leung},
    title = {Nowcasting and forecasting the potential domestic and international spread of the 2019-nCoV outbreak originating in Wuhan, China: a modelling study},
    journal = {Lancet (London, England)},
    year = {2019}
}

@article{paper18,
  author = {Vespignani, Alessandro and Tian, Huaiyu and Dye, Christopher and Lloyd-Smith, James O. and Eggo, Rosalind M. and Shrestha, Munik and Scarpino, Samuel V. and Gutierrez, Bernardo and Kraemer, Moritz U. G. and Wu, Joseph and Leung, Kathy and Leung, Gabriel M.},
  title = {Modelling {COVID-19}},
  journal = {Nature Reviews Physics},
  year = {2020},
  month = {06},
  day = {01},
  volume = {2},
  number = {6},
  pages = {279--281},
  issn = {2522-5820},
  doi = {10.1038/s42254-020-0178-4},
  url = {https://doi.org/10.1038/s42254-020-0178-4},
}

@article{paper20,
  author = {Matthew D. Holland, Alan Hastings},
  title = {Strong effect of dispersal network structure on ecological dynamics},
  journal = {Nature},
  year = {2008}
}

@article{10.1093/nsr/nwaa269,
    author = {Fan, Huawei and Kong, Ling-Wei and Wang, Xingang and Hastings, Alan and Lai, Ying-Cheng},
    title = "{Synchronization within synchronization: transients and intermittency in ecological networks}",
    journal = {National Science Review},
    year = {2020},
}

@Inbook{Allen2008,
author="Allen, Linda J. S.",
editor="Brauer, Fred
and van den Driessche, Pauline
and Wu, Jianhong",
title="An Introduction to Stochastic Epidemic Models",
bookTitle="Mathematical Epidemiology",
year="2008",
publisher="Springer Berlin Heidelberg",
address="Berlin, Heidelberg",
pages="81--130",
isbn="978-3-540-78911-6",
doi="10.1007/978-3-540-78911-6_3",
url="https://doi.org/10.1007/978-3-540-78911-6_3"
}

@article{WANG20161,
    title = {Statistical physics of vaccination},
    journal = {Physics Reports},
    volume = {664},
    pages = {1-113},
    year = {2016},
    note = {Statistical physics of vaccination},
    issn = {0370-1573},
    doi = {https://doi.org/10.1016/j.physrep.2016.10.006},
    url = {https://www.sciencedirect.com/science/article/pii/S0370157316303349},
    author = {Zhen Wang and Chris T. Bauch and Samit Bhattacharyya and Alberto d'Onofrio and Piero Manfredi and Matjaž Perc and Nicola Perra and Marcel Salathé and Dawei Zhao},
}

@article{vespignani2,
  title = {Epidemic processes in complex networks},
  author = {Pastor-Satorras, Romualdo and Castellano, Claudio and Van Mieghem, Piet and Vespignani, Alessandro},
  journal = {Rev. Mod. Phys.},
  volume = {87},
  issue = {3},
  pages = {925--979},
  numpages = {55},
  year = {2015},
  publisher = {American Physical Society},
  doi = {10.1103/RevModPhys.87.925},
  url = {https://link.aps.org/doi/10.1103/RevModPhys.87.925}
}

@article{prx2020arena,
  title = {Modeling the Spatiotemporal Epidemic Spreading of COVID-19 and the Impact of Mobility and Social Distancing Interventions},
  author = {Arenas, Alex and Cota, Wesley and G\'omez-Garde\~nes, Jes\'us and G\'omez, Sergio and Granell, Clara and Matamalas, Joan T. and Soriano-Pa\~nos, David and Steinegger, Benjamin},
  journal = {Phys. Rev. X},
  volume = {10},
  issue = {4},
  pages = {041055},
  numpages = {21},
  year = {2020},
  publisher = {American Physical Society},
  doi = {10.1103/PhysRevX.10.041055},
  url = {https://link.aps.org/doi/10.1103/PhysRevX.10.041055}
}

\end{document}